# Three-Dimensional Velocity Analysis and Particle Size Dynamics from Multi-Site RGB-Photometry of Noctilucent Clouds

Oleg S. Ugolnikov[a*], Olga Yu. Golubeva[b], Egor O. Ugolnikov[c]

[a] *Space Research Institute, Russian Academy of Sciences, Moscow, Russia*
[b] *Omsk educational institution "GDDYuT", Omsk, Russia*
[c] *V.I. Chuikov school at South-East, Moscow, Russia*

*Corresponding author e-mail: ougolnikov@gmail.com

**Abstract**
A method for measuring the altitude and particle size of noctilucent clouds, based on positioning and photometry from wide-angle three-color cameras, has been developed to determine the three velocity components, particle radius, and its derivative with respect to time for different cloud fragments. The updated method is applied to observational data of bright clouds during the summers of 2023-2025. Meridional motion of the cloud is found to be the principal factor driving the change in particle size. The effect of particle size evolution in the presence of a strong latitudinal temperature gradient is also studied.

## 1. Introduction

Noctilucent clouds (NLC) have been observed during summertime in mid- and northern latitudes since the late 19th century (Leslie, 1885; Backhouse, 1885). As suggested by Wegener (1912) and finally confirmed during the UARS mission (Hervig et al., 2001), these clouds consist of water ice in the upper mesosphere. Triangulation measurements of altitude (Jesse, 1896) yielded an average value of 82.1 km. This is several kilometers below the summer mesopause, where temperatures can be as low as 120-130 K due to upwelling air motion and adiabatic cooling. Water vapor can condense on meteoric dust particles (Rosinski & Snow, 1961) or on hydrated ions (Witt, 1969).

Growth of ice particles was modeled by Gadsden (1981) and then by Turco et al. (1982). It was shown that particles can grow to radii of up to 100 nm for $H_2O$ volume mixing ratios (VMR) of 2 ppm or more. The largest particles form the visible cloud layer at altitudes of 80-84 km. Formation of this layer requires temperatures below the frost point (145-150 K), optimally around 130 K. At even lower temperatures, ion nucleation prevails over dust nucleation, leading to a large number of small subvisible particles. The characteristic time for particle growth to a radius of 100 nm is estimated at about 10 hours. The same growth rate in radius, approximately 0.003 nm·s$^{-1}$, was found by Gadsden & Schröder (1989) for an $H_2O$ VMR 5 ppm and a temperature 140 K. The effect of particle growth duration can be examined through the phase difference between the maximum occurrence of NLC from multi-longitude observations and the temperature minimum associated with a strong 5-day planetary wave during the summer of 2025 (Ugolnikov et al., 2026). This time lag was expected to be about half the particle growth time and was found to be 3.4 ± 2.2 hours.

The particle sublimation rate in the models is defined by the descent of the largest particle to altitudes below 80 km, where the temperature is higher than the ice freezing threshold. The value of –0.01 nm·s$^{-1}$ found by Gadsden & Schröder (1989) for a temperature 150 K corresponds to sublimation time of about 3 hours. The process becomes much faster with increasing temperature.

Processes of particle nucleation and sublimation can also be influenced by horizontal, especially meridional, motion. As measured by wide-angle NLC imaging during the previous years (Ugolnikov et al., 2025), the meridional velocity is negative (southward) in most cases, reaching –50 $m \cdot s^{-1}$. During the summer, mesopause temperature decreases to the pole, and southward wind brings colder air masses to mid-latitudes (Gerding et al., 2013). As a result of this process, cloud particles can be transported to the warmer medium and then sublimate remaining at constant altitude.

To observe this process, we must trace a fixed fragment of NLC as it moves horizontally. This cannot be done by lidars, which have the constant direction of view. A decrease of mean altitude of NLC and a change of particle size during the night have been observed by lidars (Gerding et al., 2013, 2021), but at each moment in time these measurements refer to different cloud samples. Observations of a fixed cloud fragment during over a long time period are also difficult for satellite remote sensing.

Ground-based wide-angle imaging of the noctilucent cloud field provides an opportunity to track the fixed element of the cloud during its period of best visibility in twilight. If these observations are hold from two or more locations using the same calibrated RGB-cameras, then the altitude together with effective particle size can be measured (Ugolnikov, 2024). In this paper the method is extended to measure the derivative of the effective radius with respect to time, as well as the three components of fragment velocity – zonal, meridional, and vertical. Comparison of the results is applied to the several cases of bright NLC observations during recent summers, which allows identifying the primary factors governing the evolution of the ice particles in the summer mesosphere.

## 2. Observations

Noctilucent clouds were observed by a set of identical all-sky RGB cameras during the summers of 2023-2025 in central Russia, 55-57°N, 36-38°E. Spectral characteristics of the cameras are presented in (Ugolnikov, 2023), effective wavelengths for the case of NLC light scattering are 469, 529, and 578 nm. However, the analysis is performed through integration over the spectral bands rather than using the effective values. Measurements are held during twilight and night from evening to morning, with exposure times ranging from 1 to 4 seconds. Nighttime images are used to determine the camera axis position and field parameters basing on several thousand star images. Stellar photometry is also used for the determination of local atmospheric extinction in RGB bands. Data at zenith angles up to 70° are processed.

In all cases of NLC observations described here, two cameras are placed in the primary observational point, the same as in (Ugolnikov et al., 2025), while the second point is usually about 100 km northwards providing the maximal difference in scattering angles for the same NLC fragment combined with a large overlapping area. The distance between the observation points and the position angles of the baseline are given in Table 1. In the case of morning twilight of June 28, 2024, the second observation point is in the northeast direction, this is also effective for the morning conditions.

Analysis of NLC fragment motion, particle sizes, and their change in time is made based on the periods of best NLC visibility, highest contrast on the sky background, and position above the shadow of the troposphere and stratospheric ozone. These intervals are also presented in Table 1. The brightest fragments of NLC field are selected for the analysis of tiny effects related to the change in altitude and particle size during these intervals.

| Date | Distance, km | Position angle, degrees | Cameras | Time interval (UT) |
|---|---|---|---|---|
| July 3, 2023, evn | 114.7 | 6.0 | 2 + 2 | 19:34 – 19:50 |
| July 4, 2023, mrn | 114.7 | 6.0 | 2 + 2 | 23:30 – 23:50 |
| June 28, 2024, mrn | 70.3 | –56.4 | 2 + 1 | 23:15 – 23:35 |
| June 23, 2025, evn | 135.5 | 18.9 | 2 + 2 | 19:30 – 19:50 |
| June 24, 2025, mrn | 135.5 | 18.9 | 2 + 2 | 23:10 – 23:30 |
| July 3, 2025, evn | 135.5 | 18.9 | 2 + 2 | 19:30 – 19:50 |

*Table 1. Multi-site observations of NLC in 2023-2025 included to the analysis. Position angle of the line between the observation points is measured counterclockwise from the north-south line. For morning observations, the time (UT) refers to the previous date.*

### 3. Data procession

The initial stages of the processing repeat the procedure described in (Ugolnikov, 2024; Ugolnikov et al., 2025). RAW images of the sky with NLC are projected onto a surface with the fixed *a priori* altitude $H_0$ = 81.33 km taking into account of ellipsoidal model of the Earth and the small altitude of the observational points above the sea level, $h_i$. The position of the cloud element's projection onto this surface if defined by coordinates ($x$, $y$), where the axes $x$ and $y$ are directed eastward and northward, respectively. The projections of the observation points onto this surface are denoted as $x_i$, $y_i$. Following (Ugolnikov et al., 2025) in contrast to (Ugolnikov, 2024), we take the zero point on this surface above the first camera: $x_1 = y_1 = 0$. Projections of the NLC field from the image of the first camera are shown in Figure 1. The areas being processed are marked in bright, the area covered by the cameras from the second observation point is shown by the white circle.

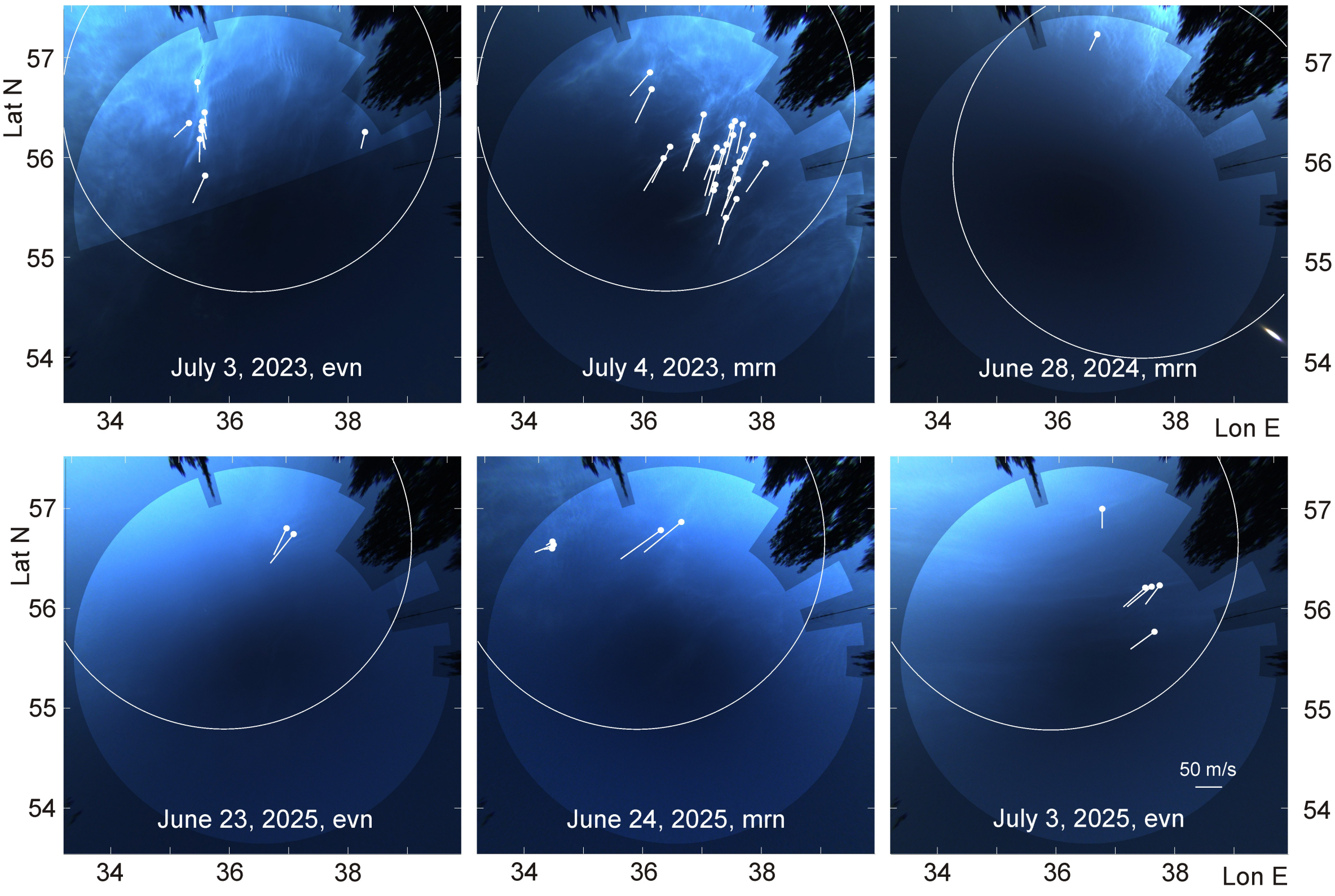

*Figure 1. Noctilucent cloud images projected onto the surface at 81.33 km above the Earth. Work are of the northern camera is shown by white circle. Optical centers and velocities of the fragments being analyzed are shown.*

Sky background subtraction procedure is applied to all images; it is described in (Ugolnikov et al., 2021) and extended to the two-dimensional case in (Ugolnikov, 2024). Following the same paper, the NLC field $J_{i1,2,3}(x, y)$ is normalized to account the effects of different path lengths through the NLC layer, lower atmospheric extinction, and scattering asymmetry:

$$J_{i\,(1.2.3)N}(x,y)=\frac{J_{i\,(1.2.3)}(x,y)\cdot\cos\gamma_i(x,y,H_0)}{S_{i(1,2,3)}(\theta_i(x,y,H_0),r_0)}\exp(\tau_{i\,(1,2,3)}\cdot AM(Z_i(x,y)).\qquad(1)$$

Here, $i$ is the camera number, $\gamma$ is the angle between the line of sight and the plumb line at the cloud element, $\tau$ is the vertical optical depth of the local atmosphere, $AM$ is the atmospheric mass corresponding to the direction from the camera to the cloud with the zenith angle $Z$. Finally, $S$ is the scattering function of an NLC particle with *a priori* radius $r_0$ = 85 nm. Indices (1, 2, 3) correspond to the RGB spectral bands. If the cloud is exactly at the altitude $H_0$ and has an effective particle size $r_0$, then the fields $J_{iN}$ will be equal for all observation sites and cameras $i$. A change of altitude causes a shift effect for different cameras, while a change in particle size leads to difference of intensities and color indices. The triangulation procedure described below is analogous to the approach in (Ugolnikov et al., 2025), but here we also compare the positions of NLC fixed by different cameras at different moments of time.

Let the cloud element have an altitude $H=H_0+\Delta H$, and it is projected onto *a priori* layer to the point ($x_N$, $y_N$). Figure 2 shows the simple case where two observational sites are aligned along the north-south direction, the cloud element lies in the same plane $x$ = 0 and has no motion along the $x$-axis. The expressions below refer to the general three-dimensional case. Following (Ugolnikov et al., 2025), we take the image of NLC obtained by the first camera in the middle of the observational period $t_0$ and define the Cartesian coordinate system with its center at the fragment being considered, axes $x_E$, $y_E$, and $z_E$ directed eastward, northward, and downward, respectively. The coordinates of the observational points in this system are written as follows (see Ugolnikov et al., 2025 for details):

$$x_{Ei}=(x_i-x_{\mathrm{N}})\frac{R+h_i}{R+H_0};$$
$$y_{Ei}=(y_i-y_{\mathrm{N}})\frac{R+h_i}{R+H_0};\qquad(2)$$
$$z_{Ei}=H_0+\Delta H-h_i+\frac{R}{2(R+H_0)^2}((x_i-x_{\mathrm{N}})^2+(y_i-y_{\mathrm{N}})^2).$$

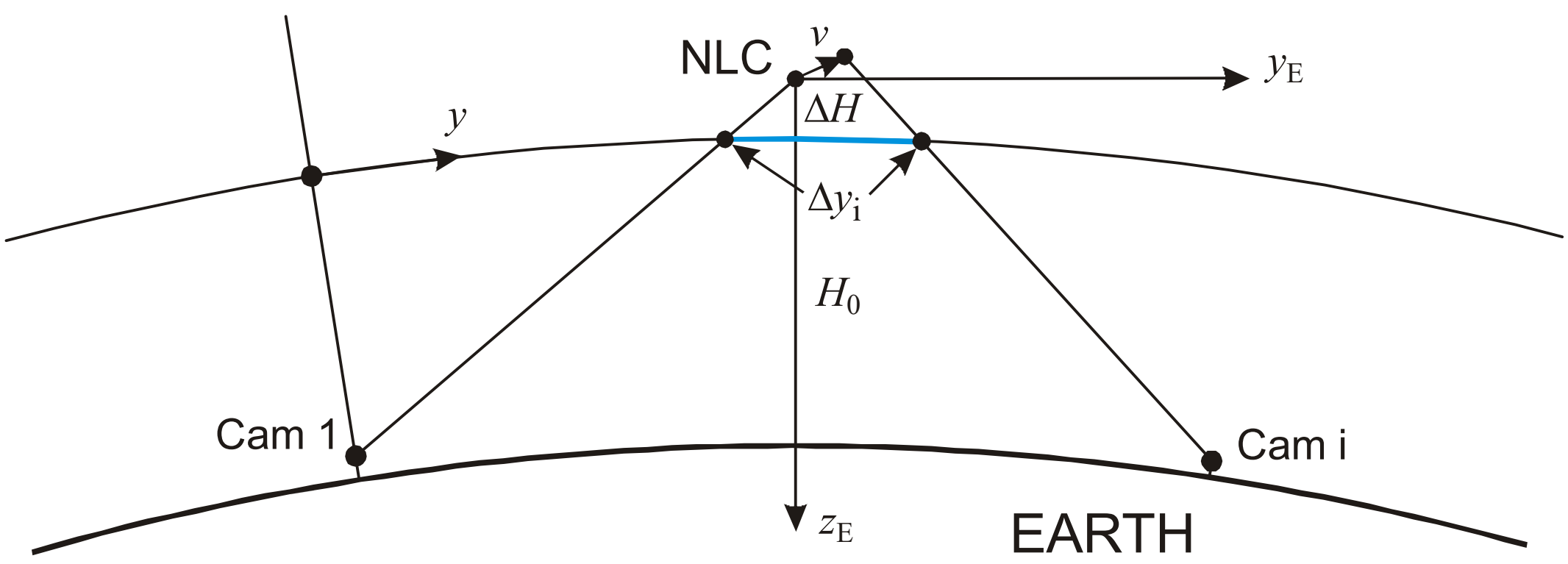

*Figure 2. Scheme of triangulation technique and coordinate systems definitions.*

Here, $R$ is the Earth's radius, $h_i$ is the small altitude of observation place. We compare the image from the first camera taken at time $t_0$ with the image of the camera $i$ taken at time $t$. The position of the fragment at this moment in the Cartesian coordinate system is given by:

$$\begin{aligned} x_{EN} &= v_X (t - t_0); \\ y_{EN} &= v_Y (t - t_0); \\ z_{EN} &= -v_Z (t - t_0). \end{aligned} \quad (3)$$

Here, $v_X$, $v_Y$, and $v_Z$ are the zonal, meridional, and vertical velocities of the cloud, respectively. We consider the vertical velocity positive if the fragment moves upward, which corresponds to a negative sign in the defined coordinate system. From these equations, we find the projection coordinates of the fragment as observed from these locations, linearizing them with respect to the small time difference $(t - t_0)$:

$$\begin{aligned} y_1(t_0) &= y_N + \frac{y_{E1}}{z_{E1}} \Delta H; \\ y_i(t) &= y_N + \frac{y_{Ei}}{z_{Ei}} \Delta H + v_Y (t - t_0) \cdot (1 - \frac{\Delta H}{z_{Ei}}) + \frac{y_{Ei} v_Z (t - t_0)}{z_{Ei}} \cdot (1 - \frac{\Delta H}{z_{Ei}}). \end{aligned} \quad (4)$$

Equations for the coordinates $x_1$ and $x_i$ are analogous. Finally, we obtain the expressions for the projections shifts:

$$\begin{aligned} \Delta x_i(t) &= \left( \frac{x_{Ei}}{z_{Ei}} - \frac{x_{E1}}{z_{E1}} \right) \cdot \Delta H + v_X (t - t_0) \cdot (1 - \frac{\Delta H}{z_{Ei}}) + \frac{x_{Ei} v_Z (t - t_0)}{z_{Ei}} \cdot (1 - \frac{\Delta H}{z_{Ei}}) + C_X; \\ \Delta y_i(t) &= \left( \frac{y_{Ei}}{z_{Ei}} - \frac{y_{E1}}{z_{E1}} \right) \cdot \Delta H + v_Y (t - t_0) \cdot (1 - \frac{\Delta H}{z_{Ei}}) + \frac{y_{Ei} v_Z (t - t_0)}{z_{Ei}} \cdot (1 - \frac{\Delta H}{z_{Ei}}) + C_Y. \end{aligned} \quad (5)$$

Calculating the values of $\Delta x$ and $\Delta y$ for all cameras and frames, we can run a least-squares procedure to find altitude correction $\Delta H$, three components of fragment velocity $v_{XYZ}$, and also free parameters $C_X$ and $C_Y$, which account for possible errors in the fragment position in the first camera field. All frames at $t \neq t_0$ for the first and second cameras in the primary observation point are also included to improve the accuracy of velocity. The first term of the equations (5) is equal to zero for the data in this point.

The first term of Equations (5) is the principal for the determination of $\Delta H$, while it is a small parameter in the second and third terms. If we take the approximate value of $\Delta H$, use it in formula (2) to find $z_{Ei}$ and the substitute it to the second and third terms in Formulae (5), these equations become linear and can be easily solved by least-squares procedure. The corrected value of $\Delta H$ obtained in this way can then be used as input for the next iteration stage. At the initial step we can assume $\Delta H = 0$; Ugolnikov et al (2025) also used a "flat approximation" with $h_i = 0$ and $z_{Ei} = H_0$ at the initial stage. 4-6 rounds of iteration are enough to find the latitude and velocity of NLC fragment.

The procedure for particle size determination originates from (Ugolnikov, 2024), where particle size was estimated for different fragments at different times. The fragmentation procedure was run independently, which caused strong variations in the derived altitude and particle size. Here, we consider the observational interval as a whole, basing on the fragmentation performed for the first camera field in the middle of the interval. Particle size determination relies on measurements of the fragment intensity in first (blue) band $J_{i1N}$ and color ratios $J_{i21N} = J_{i2N}/J_{i1N}$ (green-blue) and $J_{i31N} = J_{i3N}/J_{i1N}$ (red-blue). For each fragment of noctilucent clouds, the intensity and color indexes are found together with their rates of change over time:

$$
\begin{aligned}
J_{i1\mathrm{N}} &= J_{i1\mathrm{N}0} + \frac{dJ_{i1\mathrm{N}}}{dt}(t-t_0);\\
J_{i21\mathrm{N}} &= J_{i21\mathrm{N}0} + \frac{dJ_{i21\mathrm{N}}}{dt}(t-t_0);\\
J_{i31\mathrm{N}} &= J_{i31\mathrm{N}0} + \frac{dJ_{i31\mathrm{N}}}{dt}(t-t_0).
\end{aligned} \tag{6}
$$

Taking into account that we have already reduced the effects of local atmospheric extinction along trajectory from the cloud to the observer, the difference of angle between the viewing trajectory and NLC layer, and normalized the intensities and colors on the Mie scattering functions for $r_0 = 85$ nm, the observational parameters are compared with normalized theoretical data

$$
\begin{aligned}
S_{1\mathrm{N}}(\theta,r) &= \frac{S_1(\theta,r)}{S_1(\theta,r_0)};\\
S_{21\mathrm{N}}(\theta,r) &= \frac{S_2(\theta,r)/S_1(\theta,r)}{S_2(\theta,r_0)/S_1(\theta,r_0)};\\
S_{31\mathrm{N}}(\theta,r) &= \frac{S_3(\theta,r)/S_1(\theta,r)}{S_3(\theta,r_0)/S_1(\theta,r_0)}.
\end{aligned} \tag{7}
$$

Here, $S_{1,2,3}(\theta, r)$ are the Mie scattering functions in color bands 1, 2, 3 for scattering angle $\theta$ and particle radius $r$. Measured intensities and colors are compared with theoretical values for a fixed value of $r$ by linear regression:

$$
\begin{aligned}
J_{i1\mathrm{N}0} &= A_1 \cdot S_{1\mathrm{N}}(\theta_i,r);\\
J_{i21\mathrm{N}0} &= A_{21} \cdot S_{21\mathrm{N}}(\theta_i,r);\\
J_{i31\mathrm{N}0} &= A_{31} \cdot S_{31\mathrm{N}}(\theta_i,r).
\end{aligned} \tag{8}
$$

Here, $i$ is the camera number. Procedure is applied if the range of scattering angles $\theta_i$ is not less than 15 degrees. A wide range of scattering angles is the primary factor motivating the north-south orientation of the camera set for the particle size analysis. The particle radius $r$ is chosen by criterion of minimizing the sum of squared residuals in approximation (8). Once it is determined, we take the logarithmic derivative of the measured intensities and colors. Here we omit indices 1, 21, and 31, since the equations are similar:

$$
\frac{dJ_{i\mathrm{N}}}{J_{i\mathrm{N}}dt} = \frac{d\ln J_{i\mathrm{N}}}{dt} = \frac{d\ln A}{dt} + \frac{d\ln S_{\mathrm{N}}(\theta_i,r)}{d\theta_i}\cdot\frac{d\theta_i}{dt} + \frac{d\ln S_{\mathrm{N}}(\theta_i,r)}{dr}\cdot\frac{dr}{dt}. \tag{9}
$$

Knowing the gradient of normalized Mie scattering functions $S_{\mathrm{N}}$ with respect to the scattering angle $\theta$ and particle radius $r$, we define

$$
J'_{i\mathrm{N}} = \frac{d\ln J_{i\mathrm{N}}}{dt} - \frac{d\ln S_{\mathrm{N}}(\theta_i,r)}{d\theta_i}\cdot\frac{d\theta_i}{dt}; \quad S'_{i\mathrm{N}} = \frac{d\ln S_{\mathrm{N}}(\theta_i,r)}{dr}; \quad a = \frac{d\ln A}{dt}; \quad r' = \frac{dr}{dt}. \tag{10}
$$

The derivative value $d\theta_i/dt$ is calculated taking into account of the motion of the fragment and the change of Sun's position over time. Finally, we construct the equations for the particle nucleation/sublimation rate $r'$:

$$
\begin{aligned}
J'_{i1\mathrm{N}} &= a_1 + S'_{i1\mathrm{N}}\, r';\\
J'_{i21\mathrm{N}} &= a_{21} + S'_{i21\mathrm{N}}\, r';\\
J'_{i31\mathrm{N}} &= a_{31} + S'_{i31\mathrm{N}}\, r'.
\end{aligned} \tag{11}
$$

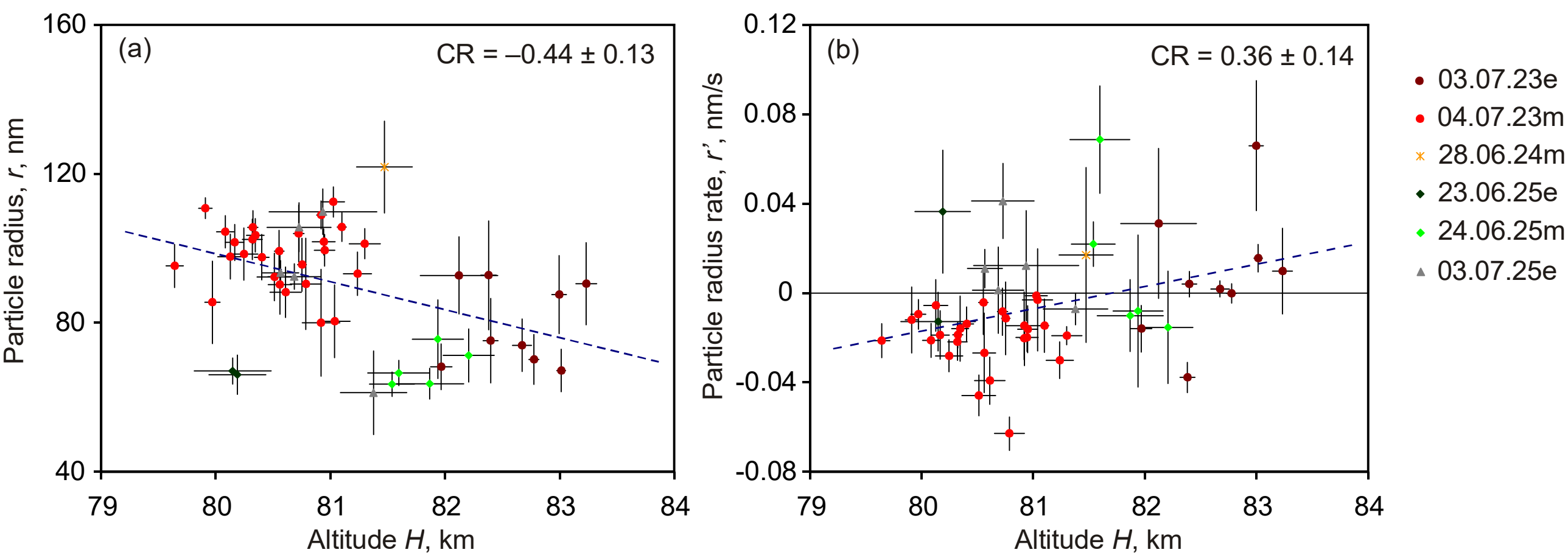

*Figure 3. Correlations of the mean altitude of the fragment with the effective particle size (a) and nucleation/sublimation rate (b).*

The parameters $a_1$, $a_{21}$, and $a_{31}$ are related to the change in brightness and color of NLC due to variations in atmospheric transparency along the tangent trajectory from the Sun and possible change in particle number density (for $a_1$). These equations are written for each camera ($i$) and can be solved by least square method if the scattering angles $\theta_i$ and values of $S$' differ significantly between observations from different sites. The results of the procedure are the values $a_1$, $a_{21}$, $a_{31}$, and $r$' – the particle nucleation/sublimation rate for a given NLC fragment.

## 4. Results

Analysis of subtle effects as the vertical motion of NLC and the change of the particle size is possible for bright fragments observed by different cameras from different locations. We present the results for all observation dates and fragments those yield good accuracy of the altitude $H_0$ (better than 0.5 km), effective particle radius $r$ (better than 15 nm) and its derivative with respect to time $r$' (better than 0.04 $\text{nm}\cdot\text{s}^{-1}$). The positions of the optical center and horizontal velocities of these fragments are shown in Figure 1. The most remarkable feature is the southward meridional velocity fixed for all observation dates and cloud fragments. This can be considered a common property of mesospheric air transport during NLC events and underscores the importance of this process for summer mesosphere conditions. The meridional velocity ranges from zero to –50-60 $\text{m}\cdot\text{s}^{-1}$. The predominant direction of the zonal velocity is westward, with the same maximal magnitude.

To study the relationship among cloud motion, altitude and physical properties, we construct the comparative diagrams of measured characteristics of the fragments. The Pearson correlation coefficient (*CR*) is shown in each diagram. Figure 3 shows the correlation between the altitude and particle radius (a) and its rate of change (b). We see that these correlations are opposite: particle radius increases at lower altitudes, while its rate becomes negative there. This reflects the descent of larger particles to the altitudes near 80 km, where they begin to sublimate.

Correlations between three velocity components and other parameters are plotted in Figures 4-6. Figure 4 shows the correlations between velocity and altitude of the fragments. We see a positive correlation, especially for the meridional velocity $v_Y$. This implies that southward transport ($v_Y < 0$) is particularly important for the clouds observations at lower altitudes about 80 km with higher temperatures and faster particle sublimation. A larger velocity helps the particles to reach lower latitudes before they evaporate and disappear. The same effect is seen in Figure 5 for correlations between the velocity and the effective particle size. Southward winds bring the large particles from polar mesosphere to the mid-latitudes.

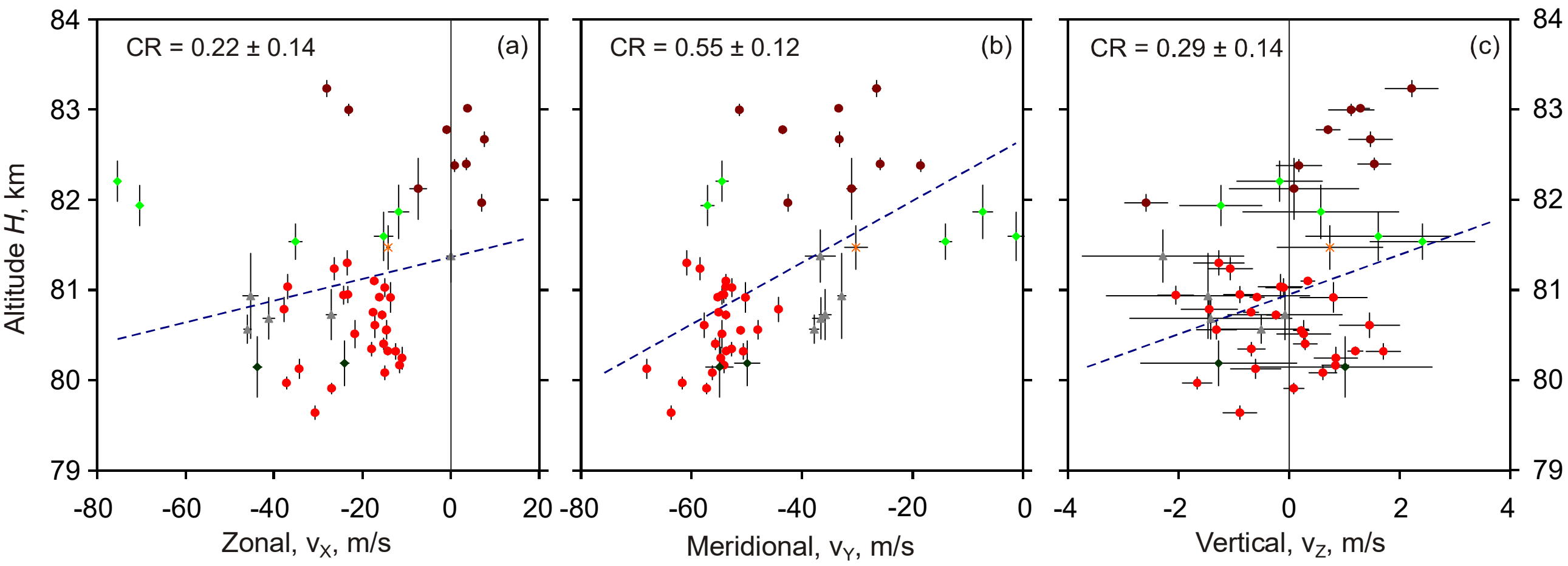

*Figure 4. Correlations of zonal (a), meridional (b), and vertical (c) velocity with mean altitude of the fragment according to observations in 2023-2025. Symbol definitions are analogous to the Figure 3.*

The vertical velocity of NLC fragments does not exceed 1-2 m·s$^{-1}$, and the average value is close to zero. However, we can see the effect of downward shift ($v_Z < 0$) for clouds at lower altitudes (a weak positive correlation between $v_Z$ and altitude). The most interesting result is the relation of velocity and particle size rate $r'$ shown in Figure 6. The small values of $r'$, comparable to their measurement errors, weaken the correlations. Nevertheless, an effect can still be seen even for small vertical velocities, indicating particle evaporation in the case of a downward shift.

Some relations are also seen for meridional velocity $v_Y$ and the particle radius rate $r'$. The overall correlation is not strong, but as shown in Figure 6b, it is evident for definite observation dates. The morning of July 4, 2023 is particularly noteworthy. Noctilucent clouds on that day were the brightest among all observations described in this paper (see Figure 1). This allowed finding the velocities and radius rates for a number of fragments. The meridional velocity was negative for all of these, reaching –60 m·s$^{-1}$ in some cases. These clouds were also the lowest among the observations (about 80 km), with effective particle radius as large as 100 nm. The radius rate $r'$ was also negative for all fragments.

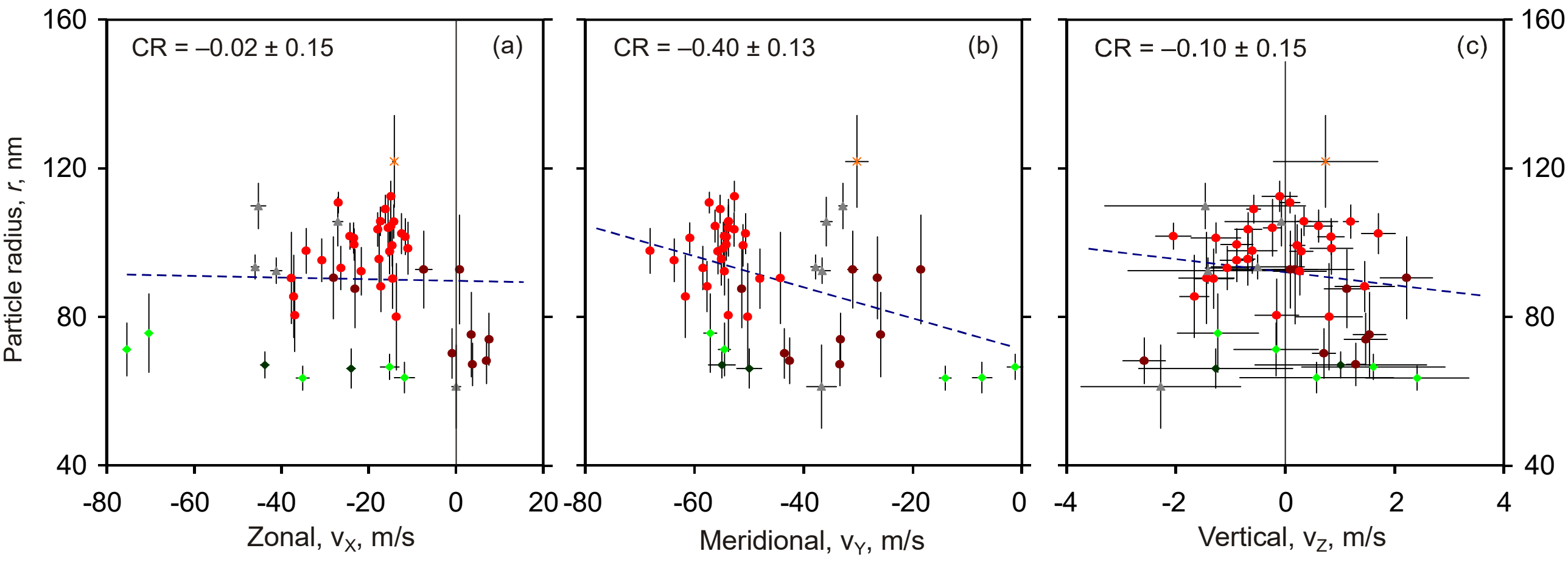

*Figure 5. Correlations of zonal (a), meridional (b), and vertical (c) velocity with mean particle radius of the fragment according to observations in 2023-2025. Symbol definitions are analogous to the Figure 3.*

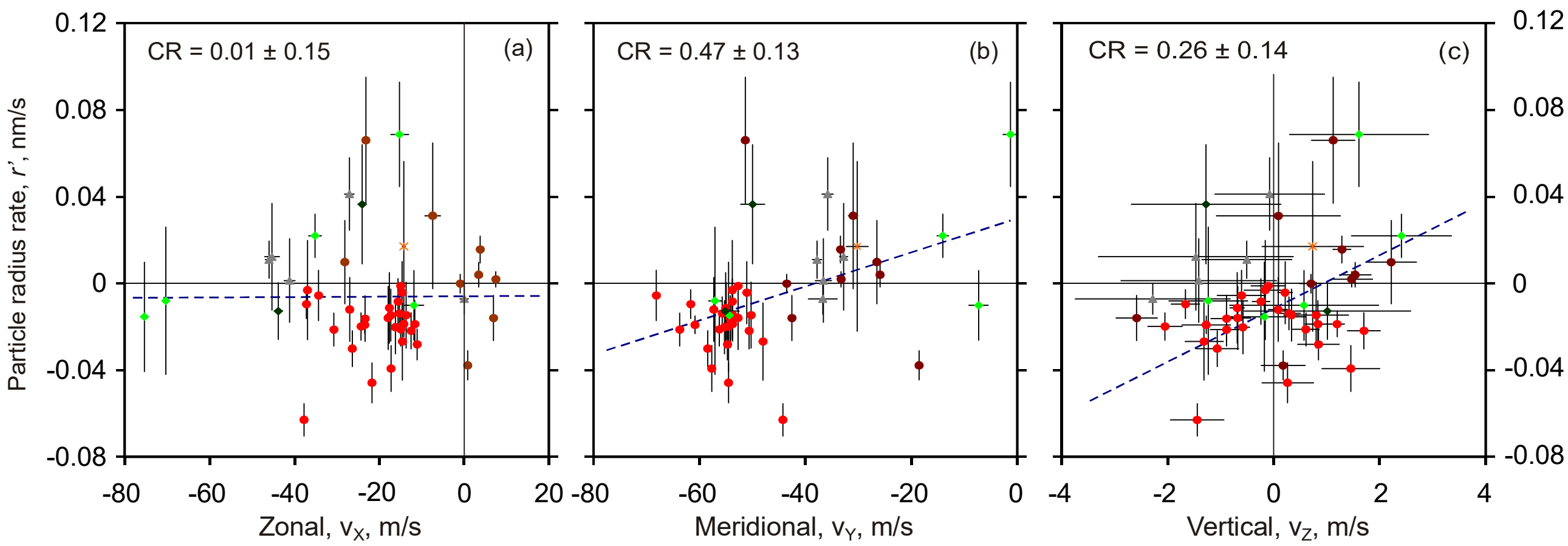

*Figure 6. Correlations of zonal (a), meridional (b), and vertical (c) velocity with nucleation/sublimation rate of the fragment according to observations in 2023-2025. Symbol definitions are analogous to the Figure 3.*

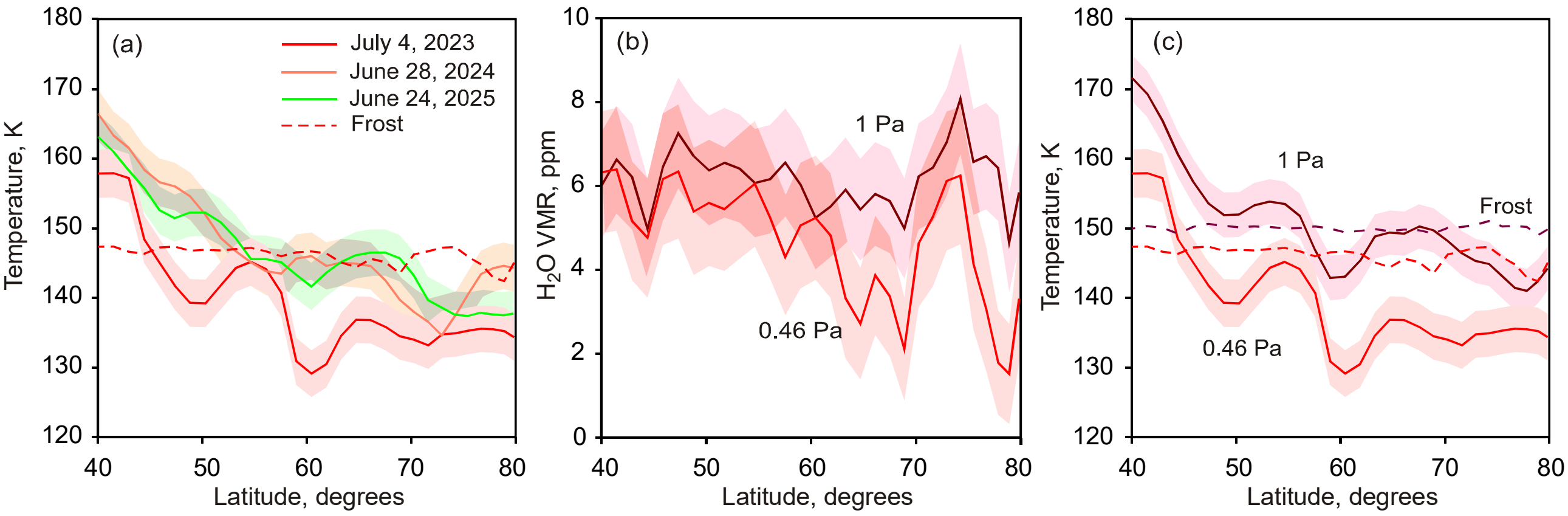

*Figure 7. Latitude dependence of EOS Aura/MLS temperature at 0.46 Pa at the nearest scan during the morning NLC observations (a); $H_2O$ VMR at 1 and 0.46 Pa in the morning of July 4, 2023 (b); temperature and ice freezing point at 1 and 0.46 Pa in the morning of July 4, 2023 (c).*

To study this phenomenon, we use the temperature dependence on latitude from nearest scan of Microwave Limb Sounder (MLS) onboard EOS Aura satellite (NASA GES DISC, 2026). The spacecraft is in a Sun-synchronous orbit, and nighttime scans are made at latitudes 55-60°N near a local solar time of 2.6 hours, which practically coincides with the time of the morning NLC observations. Latitude profiles of temperature at the 0.46 Pa level (about 82.5 km) for the morning periods of NLC observations in 2023-2025 are shown in Figure 7a. The most remarkable feature of the profile from July 4, 2023 is the rapid decrease of temperature with the latitude at 55-60°N, precisely where NLC were observed. Having formed north of 60°N at temperatures around 130K (optimal for nucleation of large particles), the clouds were transported southward into the warmer regions, causing a decrease of the mean particle radius.

July 4, 2023 was also the only observation morning for which MLS $H_2O$ profiles were available; the dependencies of $H_2O$ VMR on the latitude are shown in Figure 7b. A notable depression of $H_2O$ at 0.46 Pa and 65°N is observed, likely related to nucleation of NLC particles. Based on the $H_2O$ data, the frost point temperature can be calculated using the empirical relation of Murphy and Koop (2005); this is also shown in Figure 6a. We take into account that the mean altitude of the NLC during the morning of July 4, 2023 was lower than average, about 80 km, lying between the MLS pressure levels of 0.46 Pa (82.5 km) and 1 Pa (79 km). $H_2O$ data at the 1 Pa level are also plotted in Figure 7b, while Figure 7c shows the temperature dependence on latitude at both levels. Assuming that the NLC were located primarily between these levels, we see that the temperature crosses the ice frost point as the clouds move from 60°N to 55°N. This motion can be considered the primary factor driving particle sublimation during this period.

## 5. Discussion and conclusion

Fast motion of noctilucent clouds makes the ground-based imaging the most effective tool for their study if the physical evolution of a specific cloud fragment over time is the topic of interest. As it was shown before (Ugolnikov, 2024), RGB-photometry of the sky background with noctilucent clouds by identical cameras spread at a distance of about 100 km close to the north-south direction allows finding the altitude and effective particle radius of an NLC fragment. Here we solve a more complicated problem of finding the first derivative of these values with respect to time. Vertical velocity of NLC rarely exceeds 2 $m \cdot s^{-1}$, and the particle radius rate is about several units of $10^{-2}$ $nm \cdot s^{-1}$. These effects can be detected against the noise background only for bright cloud fragments.

Physical characteristics of the clouds are found to correlate with their horizontal velocity, especially with meridional component, which is directed southwards in the case of noctilucent clouds and can reach 60 $m \cdot s^{-1}$. This can be especially important if the temperature in the cloud layer is strongly dependent on latitude. The most remarkable case is the morning twilight of July 4, 2023, where the horizontal temperature gradient reached 0.05 $K \cdot km^{-1}$. With the velocity 60 $m \cdot s^{-1}$, this gives the warming rate about $3 \cdot 10^{-3}$ $K \cdot s^{-1}$, exceeding the warming rate due to vertical motion. The mean particle radius rate for this morning was about –0.015 $nm \cdot s^{-1}$, corresponding to a sublimation time of about 2 hours. This is found to be in good agreement with the theoretical estimations of Gadsden and Schrŏder (1989) for a temperature 150 K. The same temperature is interpolated from by EOS Aura/MLS data for this morning at an altitude 80 km, where the clouds were observed.

It is worth noting that estimation of sublimation time does not change if we assume a lognormal particle size distribution instead of a monodisperse ensemble with effective radius $r$, since both the radius and its rate with the respect to time $r'$ are reduced by the same factor (close to 2 if lognormal distribution width $\sigma = 1.4$ is assumed).

Meridional velocity and particle sublimation rate can be the basic factors defining the southern boundary of NLC observation in any given case. The relations found above point to the necessity of considering horizontal transport in future models of NLC particle evolution.

Accuracy of estimating tiny effects of change of altitude and particle size can be increased in observations at higher latitudes where NLC can be visible throughout the entire summer nights, not only during the evening and morning twilight period. However, changes in the shape of fragments restrict the range of times $(t - t_0)$ over which the correlation analysis can be performed.